\documentclass[12pt,twoside]{article}
\usepackage{fleqn,espcrc1}

\usepackage{graphicx}

\newcommand{\AmS}{{\protect\the\textfont2
  A\kern-.1667em\lower.5ex\hbox{M}\kern-.125emS}}

\title{
CP Violation from Supersymmetry in Hyperon Decays~\footnote{Talk given
at the 16th Int. Conf. on Few-Body Problem in Physics,
Taipei, Taiwan, Mar. 6-10, 2000}
}

\author{
Xiao-Gang He
        \address{
Department of Physics, National Taiwan Univeristy,
Taipei, Taiwan}}
       
\begin{document}

\maketitle

\begin{abstract}
We point out that in supersymmetric models the
direct CP violating asymmetry $A(\Lambda^0_-)$ in $\Lambda \to
\pi^- p$ can be of $O(10^{-3})$ 
in the range to be probed by the current E871 experiment.
\end{abstract}
\\

It was pointed out recently that supersymmetry can generate
flavor-changing gluonic dipole operators with sufficiently large
coefficients to dominate~\cite{1} the observed value~\cite{2}
of $(19.3\pm 2.4)\times 10^{-4}$ for
$\epsilon^\prime/\epsilon$.  
In this talk we discuss the possibility of generating large CP
violation~\cite{3} due to the same operators in the hyperon decay, $\Lambda 
\to \pi^- p$ to be measured by the E871 experiment~\cite{4}.

Experiment E871 at Fermilab is expected to reach a sensitivity 
of $2\times 10^{-4}$ for the observable 
($A(\Lambda^0_-) + A(\Xi^-_-)$) \cite{4}.  
The CP violating asymmetry $A(\Lambda^0_-)$ compares the polarization
parameter $\alpha$ for $\Lambda^0 \rightarrow p \pi^-$ 
to the corresponding parameter $\bar\alpha$ in 
$\bar\Lambda^0 \rightarrow \bar p \pi^+$ whereas $A(\Xi^-_-)$ is 
the asymmetry for the mode $\Xi^- \rightarrow 
\Lambda^0 \pi^-$. These asymmetries have a very simple form 
when one neglects the small $\Delta I =3/2$ amplitude, for 
example \cite{5},
\begin{eqnarray}
A(\Lambda^0_-) &=& {\alpha+\bar \alpha\over \alpha -\bar \alpha}
\approx -\tan(\delta_{11} - \delta_1) \sin (\phi_p -\phi_s),
\end{eqnarray}
where $\delta_{1} = 6^\circ$, $\delta_{11}=-1.1^\circ$  
are the final state $\pi N$ interaction phases for S and P wave amplitudes 
with $I =1/2$, respectively \cite{6}. $\phi_{s,p}$ are the
corresponding CP violating weak phases. Recent
calculations suggest that the strong scattering phases in 
the $\Lambda^0 \pi$ final state of the $\Xi$ decay are small \cite{7}, 
and, therefore, the current theoretical prejudice is that 
$|A(\Lambda^0_-)|$ will dominate the measurement. The standard model 
prediction for this quantity is around $3\times 10^{-5}$, albeit with large 
uncertainty \cite{5}. This suggests that a non-zero measurement 
by E871 will be an indication for new physics. 

The short distance effective Hamiltonian for the gluonic dipole
operators of interest is,
\begin{eqnarray} 
{\cal H}_{\it eff} &=& C_{g} 
{g_s\over 16\pi^2} m_s \bar d \sigma_{\mu\nu} G^{\mu\nu}_a t^a
(1+\gamma_5) s + 
\tilde{C}_{g} 
{g_s\over 16\pi^2} m_s \bar d \sigma_{\mu\nu} G^{\mu\nu}_a t^a
(1-\gamma_5) s ~+~{\rm h.c.},
\label{effh}
\end{eqnarray}
where $Tr(t^at^b) = \delta^{ab}/2$, and the Wilson coefficients 
$C_{g}$ and $\tilde{C}_{g}$ that occur in supersymmetry 
can be found in the literature \cite{8}, they are
\begin{equation}
C_{g} = (\delta^d_{12})_{LR} 
{\alpha_s \pi \over  m_{\tilde g} m_s} 
G_0(x) \eta \;,\;
\tilde{C}_{g} = (\delta^d_{12})_{RL} 
{\alpha_s \pi \over  m_{\tilde g} m_s} 
G_0(x) \eta.
\label{wilco}
\end{equation}
The parameters $\delta_{12}^{d}$ characterize the mixing in 
the mass insertion approximation \cite{8}, and 
$x=m_{\tilde g}^2/m_{\tilde q}^2$, 
with $m_{\tilde g}$, $m_{\tilde q}$ being the gluino and average  
squark masses, respectively. The loop function $G_0(x)$ is given by, 

\begin{equation}
        G_0(x)=x 
        {22-20x-2x^2+(16x-x^2+9)\log x
        \over 3 (x-1)^4}\ ,
\end{equation}
and $\eta = ({\alpha_s(m_{\tilde{g}})/ \alpha_s(m_t)})^{2/ 21}
({\alpha_s(m_t)/ \alpha_s(m_b)}
)^{2/ 23}({\alpha_s(m_b)/\alpha_s(m_c)})^{2/ 25}$ .

To calculate the weak phases we adopt the usual procedure 
of taking the real part of the amplitudes from experiment 
and of using a model for the hadronic matrix elements 
to obtain the imaginary part. Using the bag model calculations for
matrix elements 
in Ref.~\cite{3,5}, we  obtain

\begin{eqnarray}
        A(\Lambda^0_-) &=&
        \left({\alpha_s(m_{\tilde{g}}) \over \alpha_s(500~{\rm GeV})}
        \right)^{23\over 21}
        \left({500~{\rm GeV}\over m_{\tilde{g}}}\right) 
        {G_0(x)\over G_0(1)} 
         \nonumber \\
        &\times&
        \left(2.0 B_p
        [ {\rm Im}(\delta_{12}^d)_{LR}
        +{\rm Im}(\delta_{12}^d)_{RL}]
        -1.7B_s 
        [ {\rm Im}(\delta_{12}^d)_{LR}
        -{\rm Im}(\delta_{12}^d)_{RL}]
\right).
\label{cpasym}
\end{eqnarray}
We have introduced the parameters $B_s$ and $B_p$ to quantify 
the uncertainty in these matrix elements. 
To reflect these uncertainties in 
our numerical analysis we use $0.5 < B_s < 2.0$, while allowing 
$B_p$ to vary in the range $0.7 B_s < B_p < 1.3 B_s$~\cite{3}.  

In a general supersymmetric model there are also contributions 
to the imaginary parts of the Wilson coefficients of four-quark 
operators. Of these, the dominant contribution to the CP asymmetry 
in hyperon decays (within the standard model) 
is due to $O_6$ \cite{5}. We have checked numerically, that 
SUSY contributions to $C_6$ (as well as to $C_{3,4,5,7}$) are much 
smaller than those in Eq.~(\ref{cpasym}), for a parameter range 
similar to that considered in Ref.~\cite{8}.

In order to quantify $A(\Lambda^0_-)$ in supersymmetric models where 
the operators in Eq.~(\ref{effh}) have large coefficients, we 
compare Eq.~(\ref{cpasym}) with their contributions
to $\epsilon^\prime/\epsilon$ \cite{3}, 

\begin{eqnarray}
{\epsilon^\prime \over \epsilon} =
58 B_G\biggl({\alpha_s(m_{\tilde{g}}) \over \alpha_s(500~{\rm GeV})}
\biggr)^{23\over 21}
{500~{\rm GeV}\over m_{\tilde{g}}} 
{G_0(x)\over G_0(1)}\rm {Im}
\biggl( (\delta_{12}^d)_{LR}-(\delta_{12}^d)_{RL} \biggr)
{158{\rm MeV}\over m_s(m_c) + m_d(m_c)}.
\label{epspsusy}
\end{eqnarray}
In the above we have used 
the $K\rightarrow \pi\pi$ matrix element from 
a chiral quark model calculation in Ref.~\cite{9} and uses the parameter  
$B_G$ to quantify the uncertainty with 
$0.5 < B_G < 2$~\cite{10}. 
We require $\epsilon'/\epsilon$ to be 
equal to 
or less than the experimental value.

The operators in Eq.~(\ref{effh}) also contribute to $\epsilon$ 
through long distance effects. The simplest long distance  
contributions arise from $\pi^0$,~$\eta$~and~$\eta^\prime$ poles 
as noted in Ref.~\cite{11}. One has 

\begin{eqnarray}
\epsilon&=&
\biggl( {\alpha_s(m_{\tilde{g}}) \over \alpha_s(500~{\rm GeV})}
\biggr)^{23\over 21}
\biggl({500~{\rm GeV}\over m_{\tilde{g}} }\biggr)\ 
{\kappa\over 0.2}\ {G_0(x)\over G_0(1)} 6.4\  
{\rm Im}\biggl( (\delta_{12}^d)_{LR}+(\delta_{12}^d)_{RL} \biggr),
\label{epssusy}
\end{eqnarray}

The parameter $\kappa$ quantifies the contributions  
of the different poles, $\kappa=1$ corresponding to the pion pole. 
In the model of Ref.~\cite{11} 
$\kappa \sim 0.2$ whereas the contribution of the $\eta^\prime$ 
alone gives $\kappa \sim -0.9$ \cite{11}. We use 
$0.2 < |\kappa| < 1.0$ and demand that this 
long distance contribution to $\epsilon$ to be smaller 
than the experimental value $2.3\times 10^{-3}$.  

Comparing Eqs.~(\ref{cpasym})~,~(\ref{epspsusy})
and ~(\ref{epssusy}) one sees that 
$\epsilon^\prime/\epsilon$, $\epsilon$ and $A(\Lambda^0_-)$ 
are proportional to different combinations 
of the coefficients $(\delta_{12}^d)_{LR}$ and $(\delta_{12}^d)_{RL}$. 
For this reason one cannot determine the allowed range for  
$A(\Lambda^0_-)$ in term 
of $\epsilon^\prime/\epsilon$ or $\epsilon$ alone.  
We note that in general  $\epsilon'/\epsilon$ gives a stronger
constraint than $\epsilon$ if there is no cancellation among different
contributions.
In what follows, we consider three cases:
a) ${\rm Im}(\delta_{12}^d)_{RL}=0$,
b) ${\rm Im}(\delta_{12}^{d})_{LR}=0$, and 
c) ${\rm Im}(\delta_{12}^d)_{RL}= 
{\rm Im}(\delta_{12}^d)_{LR}$ for illustrations.

The case a) with $C_g$ contribution only is constrained 
by $\epsilon'/\epsilon$ to be less than $1.5\times 10^{-4}$,
and the case b) with $\tilde 
C_g$ contribution only is constrained to be less than 
$6\times 10^{-4}$.  
The case c) is particularly interesting, which is also motivated by 
theoretical considerations~\cite{12}, that there is no 
contribution to $\epsilon'/\epsilon$. 
This leads to the constraint 
$|A(\Lambda^{0}_{-})| < 7.3 \times 10^{-4} B_{p}$ by requiring SUSY
contribution to $\epsilon$ to be less than the observed value.  
Note that we allowed the range
$0.35<B_{p}<2.6$, and hence $|A(\Lambda^{0}_{-})|$ can be $O(10^{-3})$;
we cannot exclude it up to $1.9 \times 10^{-3}$. 

In summary, we have studied the supersymmetric contribution to 
CP violation in hyperon decays from gluonic dipole operators.  
We find  that the size of 
$A(\Lambda^{0}_-)$ can be of order $10^{-3}$.  
The E871 experiment will provide us with important information about
CP violation in supersymmetry. 

This work was supported in part by NSC of R.O.C. under grant number
NSC89-2112-M-002-016.
I would like to thank Murayama, Pakvasa and Valencia for collaboration on
the work presented here.

\end{document}